# Material Classification of Recyclable Containers Using 60 GHz Radar

Tommy. Albing[#1], Rikard Nelander[#2]

Acconeer AB, Sweden

[1]tommy.albing@acconeer.com, [2]rikard.nelander@acconeer.com

*Abstract* — Rather than sending used containers and materials to the landfill, recycling can help lowering the human impact on the environment. However, the process of manually sorting the mixture of incoming material can be both costly and potentially harmful to the person carrying out the task. In many cases, the manual sorting could be replaced with automation, where a container is sorted by a machine, based on a classification of the container's material. In this paper, we propose a classification algorithm, using radar data, acquired with Acconeer's A121 60GHz pulse coherent radar, for classifying liquid containers into one of the four classes metal, glass, plastic, or paper. The solution offers a cost-effective system with robust performance, able to predict the type of container with 98% accuracy.

## I. INTRODUCTION

Material classification refers to the process of determining what constitutes a given object, by processing data, acquired through some measurement of the object.

Material classification can be employed in various areas where a system benefits from the information of what a certain object is made of. One such area is recycling and waste management, where the goal is to sort the incoming material according to the material of the objects. One specific example of such a system is recycling machines for various liquid and beverage containers such as bottles and cans made from glass, plastic, metal, and paper. The goal of the system is then to sort the incoming objects according to the material, constituting them.

Acconeer's A121 60GHz pulse coherent radar[1] is a low cost, low power, and high precision radar. It can be used in a wide variety of applications and is robust to environmental factors such as ambient light conditions, dirt, and dust.

In this paper, we will demonstrate how radar can be employed in the realm of material classification, and more specifically classifying beverage containers. We propose a classification model, taking designed features, which can be deployed standalone, or as a compliment to other sensor modalities, such as camera and bar code readers.

This paper starts out discussing the proposed system and test setup, followed by a study of the radar data from various containers, from which a set of features are designed. Next, a model, taking the designed features as input, is defined for performing classification. Thereafter, the paper is concluded with result and discussion, ended by a summary of the conclusions.

## II. RELATED WORK

Related work, aiming to classify various types of solid waste, has mainly been focused on camera-based solutions, classifying using various algorithms, such as [2] and [3]. Material classification based on millimeter wave radar has been explored in [4] and [5]. However, not in the realm of waste management.

## III. METHODOLOGY

### A. Proposed system

The proposed system is depicted in Fig. 1. The radar sensor illuminates the object and measures the reflected pulse. The raw data is transferred to a host for processing. Features are extracted from the raw data. A neural network is trained using the extracted features along with the corresponding class labels. The trained model is used to classify sequences of the features.

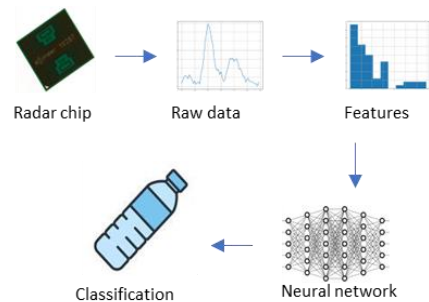

Fig. 1. A depiction of the information flow through the proposed system.

### B. Test setup

The test setup consists of several disposable containers. The objects are depicted in Fig. 2. The population was selected to reflect various sizes and shapes of each material – plastic (8), metal can (4), glass (4) paper (4). A larger number of plastic containers (8) were used as this type had a richer variety in terms of shapes and sizes.

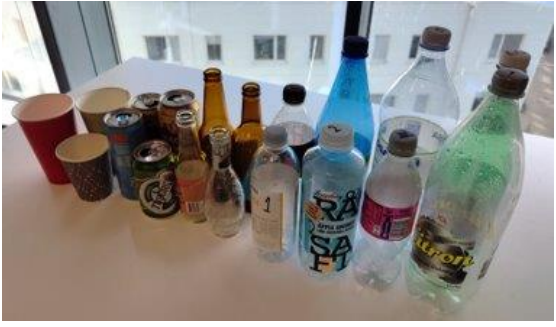

Fig. 2. A depiction of the disposable containers used in the setup.

Fig. 3. shows the setup. The container is placed on a rotating disc while the data collection is being performed. The purpose of rotating the object is to acquire a higher degree of statistically diverse data by illuminating it from various angles. The sensor is placed 250 mm from the centre of the rotating disc.

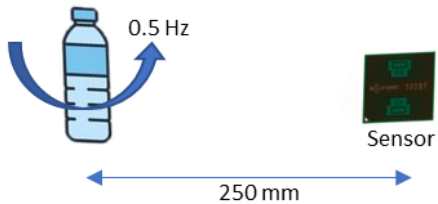

Fig. 3. A depiction of the test setup including the rotating disc and the sensor mounting(indicated by the orange square).

A frame is a set of measured radar samples, acquired from the sensor. The frame rate was set to 15 frames per second and the disc rotating at 0.5Hz, resulting in a 30 frames per revolution. One revolution of data is hereafter referred to as the *classification window*. This window constitutes the data used to produce a single prediction of the container class in the classification model.

### C. Feature Engineering

The inputs to the classification model are preprocessed features, extracted from the data in the classification window. The motivation for using preprocessed features, as opposed to providing raw data to a deep learning model, is to have a more explicit connection between the radar data and the model inputs. Also, using preprocessed features typically helps reducing the size of the classification model, while maintaining good performance.

The starting point of the feature design is the amplitudes acquired by illuminating the rotating containers with the radar. The amplitudes are depicted in Fig. 4. Each line in the plots is a realization of the measurement process. Each color represents a unique container. The graphs where two distinct peaks are present, corresponds to the case when the material has some degree of transparency, allowing the transmitted pulse to partially pass through the front of the container, reflecting off the back, resulting in a second peak.

The following key characteristics of the different container types will later be utilized when designing the features for the classification model.

- The metal can produce a high amplitude at the location corresponding to the front of the container. It has no reflection from the back of the can as no energy passes the front due to the high conductivity of the material.
- The plastic container produces significant amplitudes at both the front and the back of the container as the transmitted pulse partially passes through the front and bounces of the back of the container. The amplitude of the first and second peak are similar, with the second peak being somewhat smaller than the first peak.
- The glass container has two clear peaks, like the plastic container. The ratio of the peak amplitudes is greater, compared to the plastic containers. The main and secondary peak amplitudes as a large variance, compared to the plastic case.
- The paper containers share characteristics with both the plastic and glass containers, but with overall lower amplitude.

Based on these observations, the following features were selected as inputs to the classification model.

- Main peak mean amplitude – Measured amplitude, reflected from front of container.
- Secondary peak mean amplitude – Measured amplitude, reflected of back of container.
- Peak amplitude ratio – The ratio between the main and secondary peak mean amplitudes.
- Main peak amplitude variance – The variance of the amplitudes, acquired while rotating the container, at the main peak.
- Secondary peak amplitude variance - The variance of the amplitudes, acquired while rotating the container, at

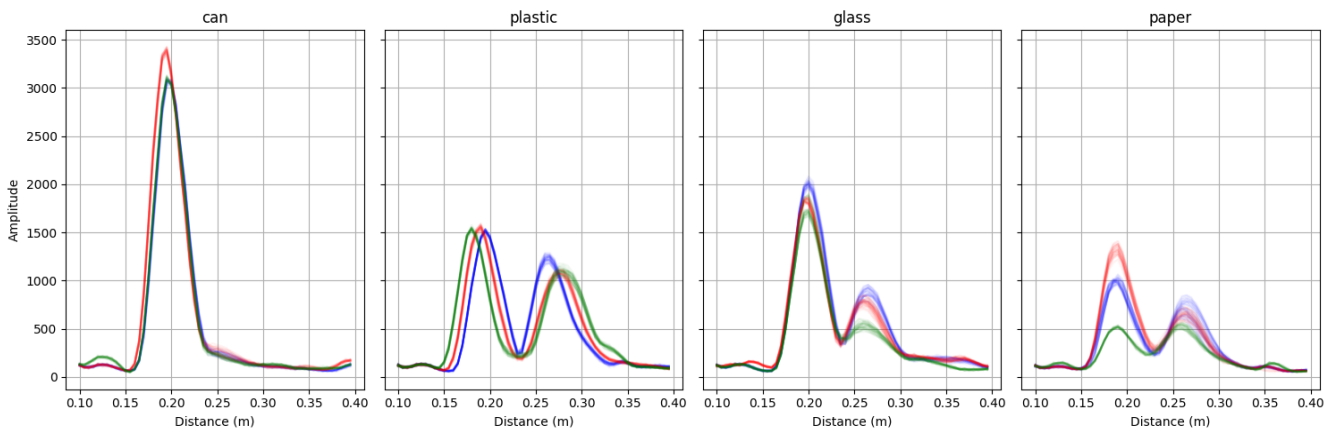

Fig. 4. The measured amplitude of for the four different container types. Each color represents a new container of different size and shape of the given type.

the secondary peak.
- Peak variance ratio – The ratio between the variance of the main and secondary peak amplitudes.

These six features, along with the class label, were extracted from each classification window and used for training the classification model.

### D. Model selection

The classification model is implemented using Google's Tensorflow [6].

In addition to the four material classes(metal can, plastic, glass and paper), a fifth class, referred to as *empty,* is added, reflecting a setup with no container present in the scene.

The selected model consists of two densely connected layers, according to the following configuration.
- Normalization layer with input size 6(number of features).
- Densely connected layer with 50 neurons and relu activation function.
- Dropout layer with dropout rate of 0.1.
- Densely connected layer with 40 neurons and relu activation function.
- Dropout layer with dropout rate of 0.1.
- Densely connected layer with 10 neurons and relu activation function.
- Densely connected layer with 5 neurons (representing the five available classes) and softmax activation function.

The model is trained for 100 epochs using a batch size of 32 samples. The test and training set is shuffled and split into two 30/70% datasets.

The model was trained using class weights to counteract the model favouring a certain type of container due to the varying number of test objects of the different classes.

## IV. RESULTS AND DISCUSSION

The resulting model achieves an accuracy of 98.0% with the following confusion matrix, evaluated using the test data set.

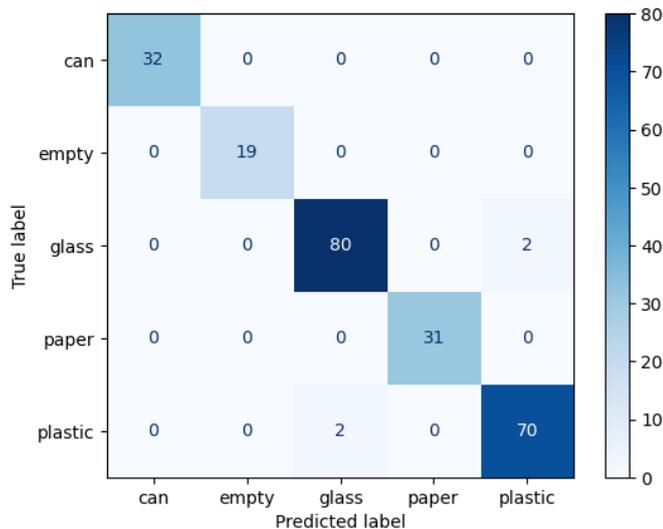

Fig. 5. The confusion matrix. Illustrating the predicted vs the actual class.

As can be seen from Fig. 5., there are 4 cases of misclassification of the plastic and glass classes. This could be due to the fact that the selected features of the two materials and shapes are somewhat similar.

One possible alternative to further improve the result and lower the miss-classifications is to extend the classification window to include a longer time series. However, this would increase the required measurement time.

Another aspect that could be evaluated it so use an array of sensors, or alternatively move the container along the sensors position while rotating, to get an even more diverse statistical representation of the container.

The rotation of the container could also be replaced with an integration utilizing multiple sensors, mounted in a ring-formation around the container.

The model structure was selected by iterating different configurations of densely connected networks. Other classification models could be evaluated to achieve higher accuracy.

The data acquisition was performed using empty bottles. Residual liquid in the container could potentially degrade the performance of the algorithm but could be alleviated by scanning the bottles in an upright position, or using some screening method where non-empty bottles would be sorted out.

All the containers used in this article had a circular cross section. A square, or other geometry, could potentially require modifications to the test setup or how the scanning is performance. For instance, scanning the container along the upright position rather than around it, could be a better performing strategy.

## V. CONCLUSION

This article presents an efficient method for classifying the material of various containers. The methodology could be applied in a range of areas where the material type is of interest to the overall system.

### ACKNOWLEDGMENT

This work was supported by the Core Industry IT Convergence Program (20016243, 336 Development of AI Embedded Radar Sensor and System Operation Platform with Object Recognition and Counting Technologies for Manufacturing Logistics AGVs), funded by the Ministry of Trade, 338 Industry & Energy (MOTIE, Korea).